\documentclass[A4, 10 pt, conference]{ieeeconf}  

\IEEEoverridecommandlockouts                              

\usepackage[latin1]{inputenc} 
\usepackage{fancyhdr}
\usepackage[hang]{caption2}
\usepackage[english]{babel}
\usepackage{color}

\usepackage{epsfig}
\usepackage{graphicx}
\usepackage[normalsize,it,hang]{subfigure}
\usepackage{amsfonts,amsmath,bm,hhline}
\usepackage{xcolor}
\newtheorem{remark}{Remark}

\title{\LARGE \bf
Easily implementable time series forecasting techniques for \\ resource provisioning in cloud computing
}

\author{Michel Fliess$^{1,3}$, Cédric Join$^{2,3}$, Maria Bekcheva$^{4,5}$, Alireza Moradi$^{4}$, Hugues Mounier$^{5}$
\thanks{$^{1}$LIX (CNRS, UMR 7161), \'Ecole polytechnique, 91128 Palaiseau, France, Michel.Fliess@polytechnique.edu}
\thanks{$^{2}$CRAN (CNRS, UMR 7039), Université de Lorraine, BP 239, 54506 Vand{\oe}uvre-lès-Nancy, France, cedric.join@univ-lorraine.fr}
\thanks{$^{3}$AL.I.E.N. (ALg\`{e}bre pour Identification \& Estimation Num\'{e}riques), 7 rue Maurice Barr\`{e}s, 54330 V\'{e}zelise, France, \newline \{michel.fliess, cedric.join\}@alien-sas.com}
\thanks{$^{4}$Inagral, 128 rue  de la Bo\'{e}tie, 75008 Paris, France, \newline  \{maria, alireza\}@inagral.com} 
\thanks{$^{5}$Laboratoire des Signaux et Syst\`emes (L2S),  Universit\'e Paris-Sud-CNRS-CentraleSup\'elec, Universit\'e Paris-Saclay, 
91192 Gif-sur-Yvette, France, \{maria.bekcheva, hugues.mounier\}@l2s.centralesupelec.fr}
}

\begin{document}

\maketitle
\thispagestyle{empty}
\pagestyle{empty}

\begin{abstract}
Workload predictions in cloud computing is obviously an important topic. Most of the existing publications employ various time series techniques, that might be difficult to implement. We suggest here another route, which has already been successfully used in financial engineering and photovoltaic energy. No mathematical modeling and machine learning procedures are needed. Our computer simulations via realistic data, which are quite convincing, show that a setting mixing algebraic estimation techniques and the daily seasonality behaves much better. An application to the computing resource allocation, via virtual machines, is sketched out.

\textit{Key words}--- Cloud computing, computing resources, virtual machines, forecasting, time series, nonstandard analysis, trend, quick fluctuation, seasonality, estimation, machine learning.
\end{abstract}

\section{Introduction}
The fast development of cloud computing (see, \textit{e.g.}, \cite{berkeley,marinescu}) is due to its tremendous benefits, which might be summarized by its \emph{pay-as-you-go} (\emph{PAYG}) feature. Providing adequate \emph{Quality-of-Service} ({\emph{QoS}) and  pricing necessitates obviously a good prediction of the workloads for a better resource provisioning. Several papers have already been written on this topic (see, \textit{e.g.}, \cite{compar,pricing,iran,balaj,arima,gu,han,huang,islam,kaur,kim,kumar,roum,vazquez,ye}, and the references therein). Various time series techniques prevail, especially those stemming from econometrics \cite{melard,tsay}, artificial neural networks and deep learning \cite{la,yan}, or a mixture of both \cite{zhang}. They all need cumbersome mathematical modeling and/or machine learning procedures.  

This communication suggests that another rather efficient approach to workload forecasting exists where the computer burden is much less demanding. This new viewpoint  on time series, that was introduced ten years ago in financial  engineering \cite{perp,agadir}, rests on a profound result \cite{cartier} which was obtained via the language of \emph{nonstandard analysis} \cite{robinson}. It has been successfully employed for the prediction of traffic flow on highways \cite{troyes} and mainly of photovoltaic energy \cite{solar}. It is worth mentioning that the signals corresponding to the workloads here and to the solar power in \cite{solar} are quite similar including their sampling. 

Contrarily to many publications, 
\begin{itemize}
\item deterministic and probabilistic/statistical modelings become useless and therefore also parameter identification and/or machine learning,
\item we are not trying to employ time series for taking into account quick oscillations of the workload. Realistic computer experiments \cite{iste} show that \emph{intelligent proportional controllers}, or \emph{iP}s, which are derived from the \emph{model-free control} setting \cite{csm}, achieve this \emph{elasticity} property (see, \textit{e.g.}, \cite{al,cou}) very well. 
\end{itemize}
Lack of space prevents us to detail the computer implementations. Note however that this implementation of our algebraic tools has already been fruitfully completed not only with respect to time series (see, \textit{e.g.}, \cite{solar}), but also in control engineering and signal processing (see, \textit{e.g.}, \cite{andrea,iste,beltran,ramp,sira1,sira2}).

Our paper is organized as follows. Our viewpoint on time series is developed in Section \ref{ts}, as well as three prediction techniques. Computer simulations are displayed and discussed in Section \ref{simu}. An application to the computing 
resource allocation, via virtual machines, is sketched out in Section \ref{virtual}. Section \ref{conclusion} presents some concluding remarks.

\section{Time series} \label{ts}
\subsection{Time series and nonstandard analysis}
\subsubsection{Nonstandard analysis}
Nonstandard analysis was invented by Robinson \cite{robinson} almost sixty years ago in order to give a rigorous definition of ``infinitely small'' and ``infinitely large'' numbers. For more readable initiations, see, \textit{e.g.}, \cite{diener1,diener2}. Let us emphasize that  this unconventional achievement has also been employed in applied sciences (see, \textit{e.g.}, \cite{harthong}).  
\subsubsection{A nonstandard definition of time series}
Take the time interval $[0, 1]$. Introduce as often in
nonstandard analysis the infinitesimal sampling
\begin{equation*}\label{sampling}
T = \{ 0 = t_0 < t_1 < \dots < t_\nu = 1 \}
\end{equation*}
where $t_{i+1} - t_{i}$, $0 \leq i < \nu$, is {\em infinitesimal},
{\it i.e.}, ``very small.'' A time series $X$ is a function
$T \rightarrow \mathbb{R}$.
\begin{remark} \label{rem}
In practice a time lapse of  $1$ minute should be viewed as quite small when compared to $1$ day. 
\end{remark}
\subsubsection{Quick fluctuations}
A time series ${\frak{X}}: T \rightarrow \mathbb{R}$
is said to be {\em quickly fluctuating}, or {\em oscillating}, around $0$ \cite{cartier}, if,
and only if, its integral $\int_A {\frak{X}} dm$ \cite{cartier} is
infinitesimal for any \emph{appreciable} interval $A$, \textit{i.e.}, an interval which is neither infinitely small nor infinitely large. 
\begin{remark}
Let us emphasize that the probabilistic and/or statistical nature of those quick fluctuations do not play any role.
\end{remark}

\subsubsection{The Cartier-Perrin theorem}
The Cartier-Perrin theorem \cite{cartier} states\footnote{The presentation in \cite{lobry} is less technical. We highly recommend this paper. It also includes a fruitful discussion on nonstandard analysis.} for a time series $X: {\mathfrak{T}} \rightarrow \mathbb{R}$ satisfying a rather weak integrability assumption the following additive decomposition
\begin{equation}\label{decomposition}
\boxed{X(t) = E(X)(t) + X_{\tiny{\rm fluctuat}}(t)}
\end{equation}
where
\begin{itemize}
\item the \emph{mean} $E(X)(t)$ is {\em Lebesgue integrable},
\item $X_{\tiny{\rm fluctuat}}(t)$ is quickly fluctuating.
\end{itemize}
The decomposition \eqref{decomposition} is unique up to an additive
infinitesimal quantity.
\begin{remark}
Replace the word ``mean'' by \emph{trend}, which is perhaps more popular, in financial engineering for instance. Set therefore
$$E(X)(t) = X_{\rm trend}(t)$$
Note however that the meaning of ``trend'' in the time series literature is most often quite different (see, \textit{e.g.},  \cite{tsay}).
\end{remark}
\begin{remark}
Our mean or trend should be understood as being quite close to the familiar notion of \emph{moving average} \cite{melard,tsay}.
\end{remark}

\subsection{Three forecasting techniques}\label{predict}
According to the Cartier-Perrin theorem \eqref{decomposition} it only makes sense to forecast $E(X)(t)$, \textit{i.e.}, the mean or the trend. 

\subsubsection{Scaled persistence and seasonality\protect\footnote{\emph{Persistence} and \emph{scaled}, or \emph{smart}, \emph{persistence} are quite often discussed elsewhere in the literature, essentially perhaps in meteorology (see, \textit{e.g.}, \cite{voyant2,voyant}) and climatology (see, \textit{e.g.}, \cite{mudel}).}}\label{scaled}
The \emph{persistence} assumption reads in our context
\begin{equation}\label{persist}
 \widehat{X_{\rm trend}}(t + h) = X_{\rm trend}(t) 
\end{equation}
where $h > 0$. Obviously Equation \eqref{persist} will too often yield poor predictions. That is why we introduce \emph{scaled persistency} via the most classic notion of  \emph{seasonality} in the time series literature \cite{melard,tsay}. Our data, where the sampling period is equal to $1$ minute, exhibit here a self-evident daily pattern. Equation \eqref{persist}  should then be replaced by 
\begin{equation}\label{scal}
 \widehat{X_{\rm trend}}(t + h) = \frac{X_{\rm trend}(t - 1440 + h)}{X_{\rm trend}(t - 1440)} X_{\rm trend}(t )
\end{equation}
where 
\begin{itemize}
\item $0 < h \leq 60 {\rm min}$, 
\item $\frac{X_{\rm trend}(t - 1440 + h)}{X_{\rm trend}(t - 1440)}$ is a correcting multiplicative term corresponding to the daily seasonality, 
\item $1440 = 60 \times 24$ is the number of minutes during a single day.
\end{itemize}

\subsubsection{Forecasting via algebraic estimation techniques\protect\footnote{For the estimation techniques, see also \cite{easy,sira1}, and \cite{mbo} for more mathematical details.}}\label{alg}
Start with a polynomial time function 
$$p_1 (t) = a_0 + a_1 t, \quad t \geq 0, \quad a_0, a_1 \in \mathbb{R},$$ of degree $1$. Rewrite
thanks to classic operational calculus (see, \textit{e.g.},
\cite{yosida}) $p_1$ as 
$$P_1 = \frac{a_0}{s} + \frac{a_1}{s^2}$$ 
Multiply both sides by $s^2$:
\begin{equation}\label{1}
s^2 P_1 = a_0 s + a_1
\end{equation}
Take the derivative of both sides with respect to $s$, which
corresponds in the time domain to the multiplication by $- t$:
\begin{equation}\label{2}
s^2 \frac{d P_1}{ds} + 2s P_1 = a_0
\end{equation}
The coefficients $a_0, a_1$ are obtained via the triangular system
of equations (\ref{1})-(\ref{2}). We get rid of the time
derivatives, \textit{i.e.}, of $s P_1$, $s^2 P_1$, and $s^2 \frac{d
P_1}{ds}$, by multiplying both sides of Equations
(\ref{1})-(\ref{2}) by $s^{ - n}$, $n \geq 2$. The corresponding
iterated time integrals are low pass filters. They attenuate the
corrupting noises, which are viewed as highly fluctuating phenomena. A quite short time window is sufficient for
obtaining accurate values of $a_0$, $a_1$.

The extension to polynomials of higher degree is obvious, and therefore also to truncated Taylor expansions.

\begin{remark}
In practice, the above integrals are of course replaced by straightforward linear digital filters.
\end{remark}

Assume that the following rather weak assumption holds true: the mean $E(X(t))$ may be associated to a differentiable time function $[0, 1] \rightarrow \mathbb{R}$. Then, on a short time lapse, $E(X(t))$ is
well approximated by a polynomial function of degree $1$. The above calculations yield via sliding time windows numerical estimates $X_{\rm trend}(t)$ and $\dot{X}_{\rm trend}(t)$ of the trend  and of its derivative. Causality is taken into account via backward time calculations. 
In this setting (\cite{perp,agadir}), forecasting the time series $X(t)$ boils down to an extrapolation of its mean $E(X(t))$. If $h > 0$ is not too ``large'', \textit{i.e.}, a few minutes in our context, a first order Taylor expansion yields the following estimate at time $t + h$
\begin{equation}\label{PredS}
\widehat{X_{\rm trend}}(t + h) = X_{\rm trend}(t) + {\dot {X}_{\rm trend}}(t) h 
\end{equation}

\subsubsection{Algebraic estimation and seasonality}\label{mixte}
Replace Equation \eqref{PredS} by
\begin{equation}\label{Mix}
\widehat{X_{\rm trend}}(t + h) = X_{\rm trend}(t) + {\dot {X}_{\rm trend}}(t - 1440 + h) h 
\end{equation}
where ${\dot {X}_{\rm trend}}(t - 1440 + h)$ is the derivative $1$ day backwards. With such a choice the derivative estimation needs not to be causal and becomes more precise and much easier to compute \cite{mboup}.

\section{Computer simulations}\label{simu}
\subsection{Data}\label{A}
The time series ${\cal X}(t)$ was provided by the Company Inagral  to whom two authors, M. Bekcheva and A. Moradi, belong:
\begin{itemize}
\item it was recorded during a time lapse $\Delta$ equal to $10$ days, with a sampling period of $1$ minute,
\item it represents the sum of processing times (CPU times) of the incoming user requests on a production Web Service, with a sampling rate of $1$ minute.
\end{itemize}
Replace ${\cal X}(t)$ by (see Figure \ref{trends}) 
$$y(t) = \frac{{\cal X}(t)}{{\rm max}_{\tau \in \Delta} {\cal X}(\tau)}$$
Thus $0 \leq y(t) \leq 1$, $\forall t \in \Delta$. This normalization procedure hides any sensitive information. The practical meaning of our computer simulations should nevertheless remain clear.


\subsection{Comparison between the three different techniques}
Predictions stemming from the three Equations \eqref{scal}, \eqref{PredS}, \eqref{Mix} are now compared. In order to derive a sound comparability procedure let us assume the following property:
Consider the three time series associated to the three forecasting techniques. In each case the quick fluctuations around the trend may be viewed \cite{noise} as a \emph{noise} like in engineering. It therefore yields a \emph{signal-to-noise ratio} (\emph{SNR}) \cite{proakis}. The three corresponding SNRs are assumed to be approximatively equal.
 
Three time horizons are considered: $5$, $30$, $60$ minutes. The results are reported in the table below, where the first (resp. second, third) column corresponds to Formula \eqref{scal} (resp.  \eqref{PredS}, \eqref{Mix}):
\begin{table}[h!]
\begin{center}
\caption{$\sum Error^2$}\label{tb}
\begin{tabular}{cccc}
Prediction horizons & Pe & Al [gain in \%] & Mi [gain in \%] \\\hline
$t+5{\rm min}$ & 15.12 & 12.47 [21.21\%] & 10.49  [44.17\%]\\
$t+30{\rm min}$ & 32.23 & 65.68 [-50.862\%] &27.89  [15.56\%] \\ 
$t+60{\rm min}$ & 53.49 & 153.79 [-65.21\%] & 49.04 [9.08\%] \\ \hline
\end{tabular}
\end{center}
\end{table} \\
The superiority of the approach from Section \ref{mixte}, which is mixing the daily seasonality with our algebraic calculations, is indubitable. See
Figure \ref{trends} for a view of the data and of the various trends according to the forecast horizons. Figure \ref{Z1} (resp. \ref{Z2}, \ref{Z3}) displays forecasts according to Section \ref{scaled} (resp. \ref{alg}, \ref{mixte}).

\begin{figure*}[!ht]
\centering%
\subfigure[\footnotesize View]
{\epsfig{figure=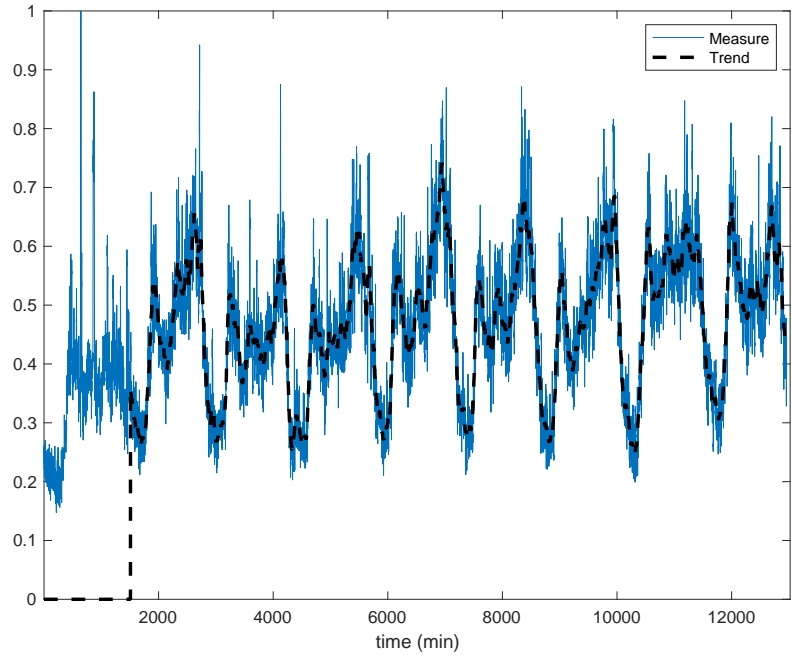,width=0.72\textwidth}}
\\
\subfigure[\footnotesize Zoom1]
{\epsfig{figure=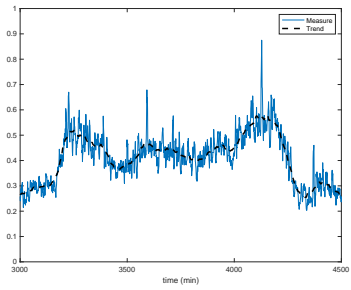,width=0.31\textwidth}}
\subfigure[\footnotesize Zoom2]
{\epsfig{figure=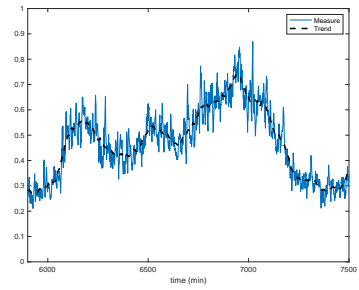,width=0.31\textwidth}}
\subfigure[\footnotesize Zoom3]
{\epsfig{figure=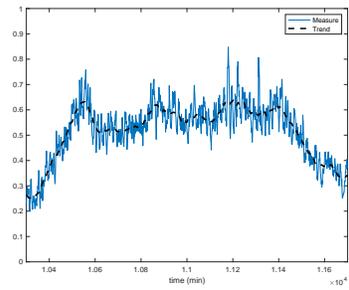,width=0.31\textwidth}}
\caption{Time series and its trend}\label{trends}
\end{figure*}


\begin{figure*}[!ht]
\centering
\subfigure[\footnotesize Horizon: 5min]
{\epsfig{figure=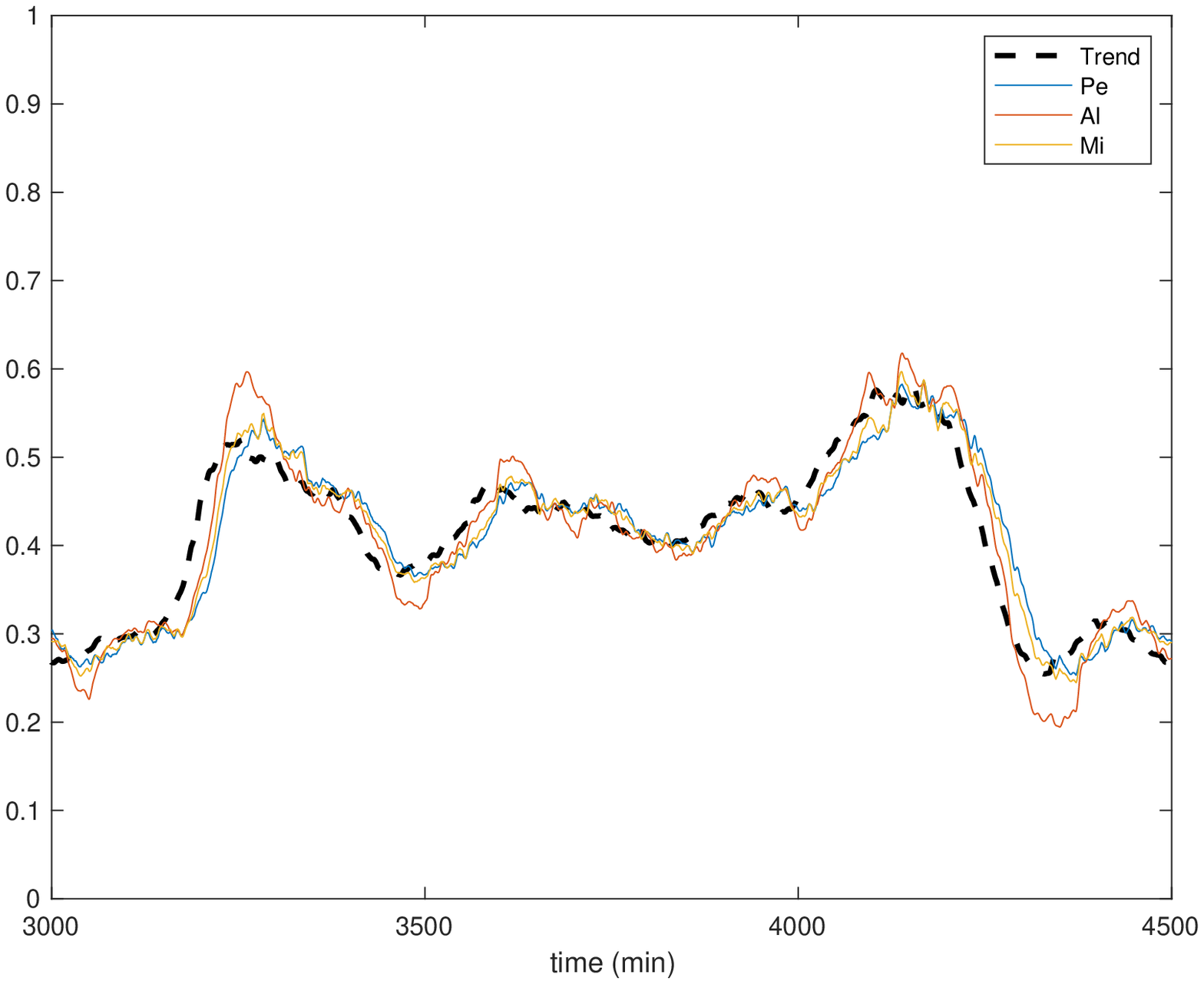,width=0.31\textwidth}}
\subfigure[\footnotesize Horizon: 30min]
{\epsfig{figure=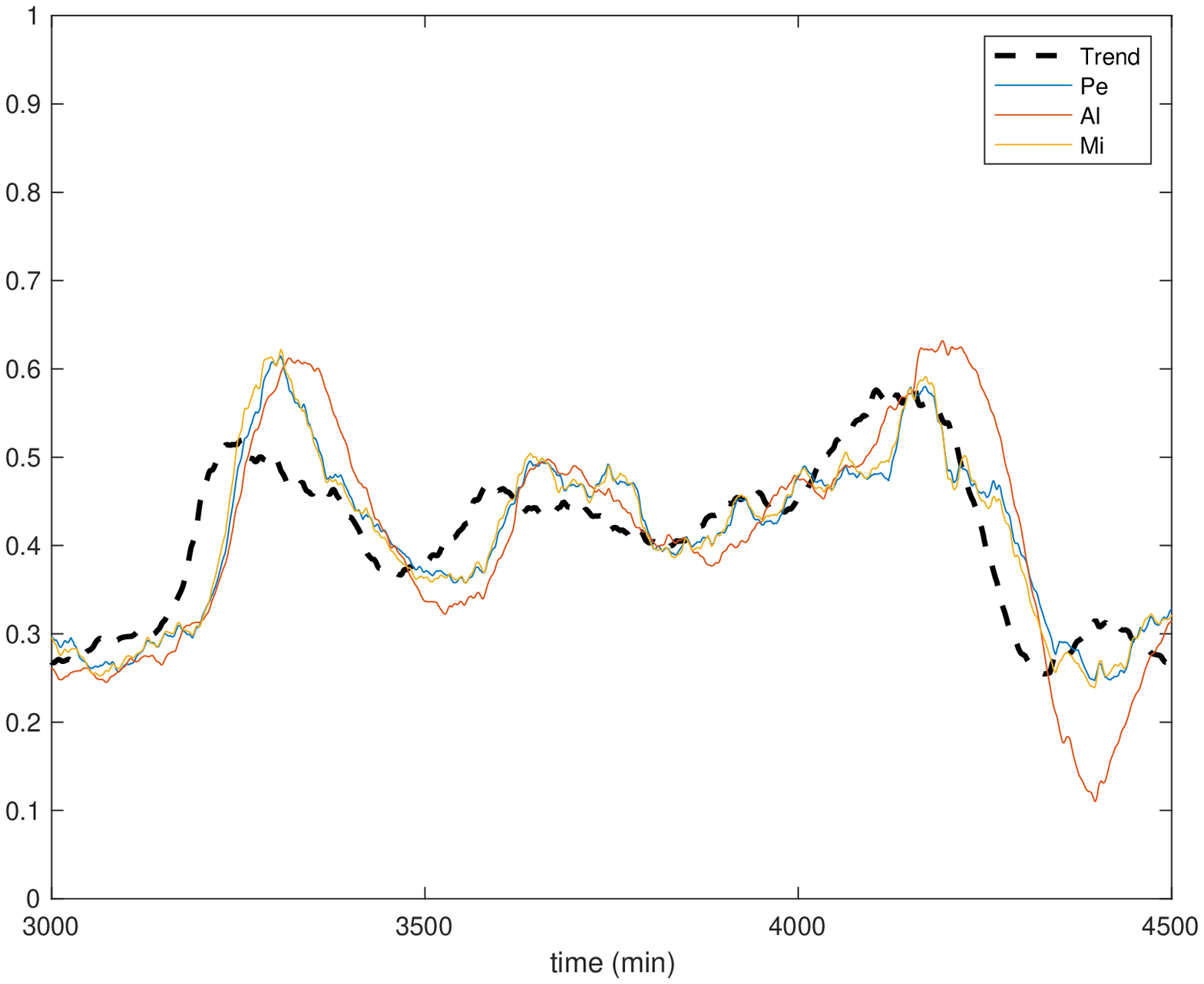,width=0.31\textwidth}}
\subfigure[\footnotesize Horizon: 60min]
{\epsfig{figure=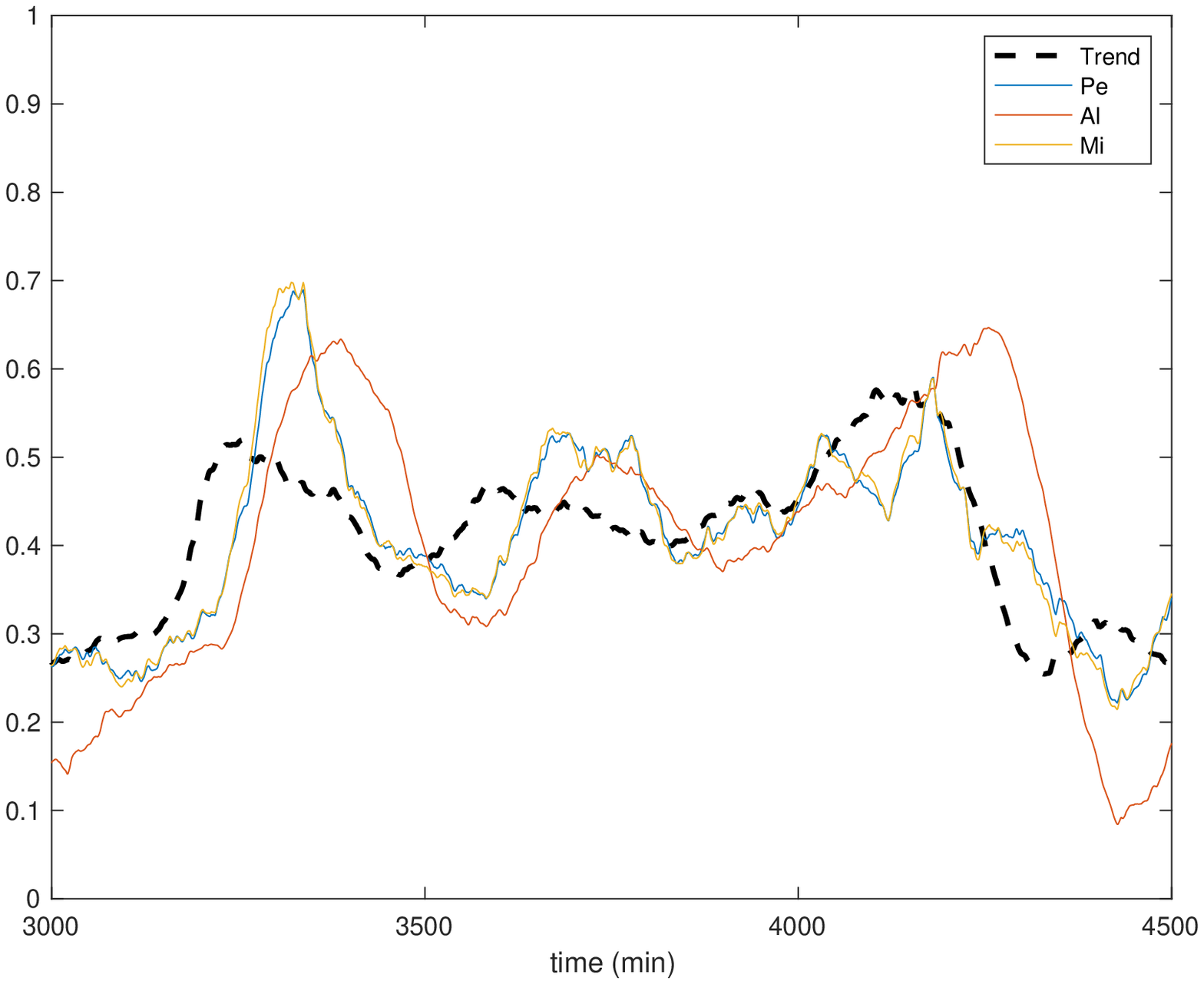,width=0.31\textwidth}}
\caption{Zoom1: Scaled persistence (\textcolor{blue}{\textbf{---}}), Algebraic techniques (\textcolor{red}{\textbf{---}}) and Algebraic techniques \& seasonality (\textcolor{yellow}{\textbf{---}}).}\label{Z1}
\end{figure*}


\begin{figure*}[!ht]
\centering
\subfigure[\footnotesize Horizon: 5min]
{\epsfig{figure=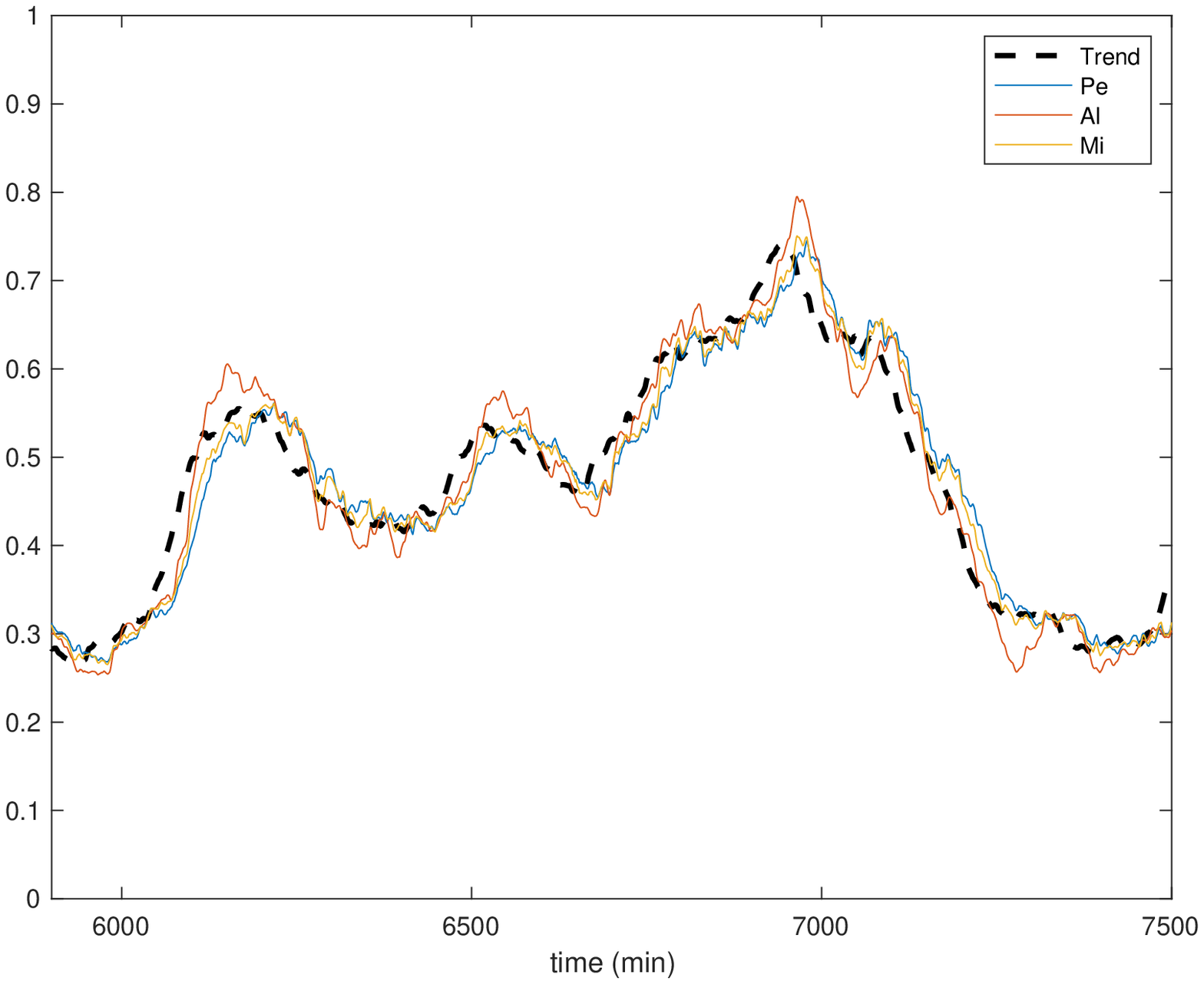,width=0.31\textwidth}}
\subfigure[\footnotesize Horizon: 30min]
{\epsfig{figure=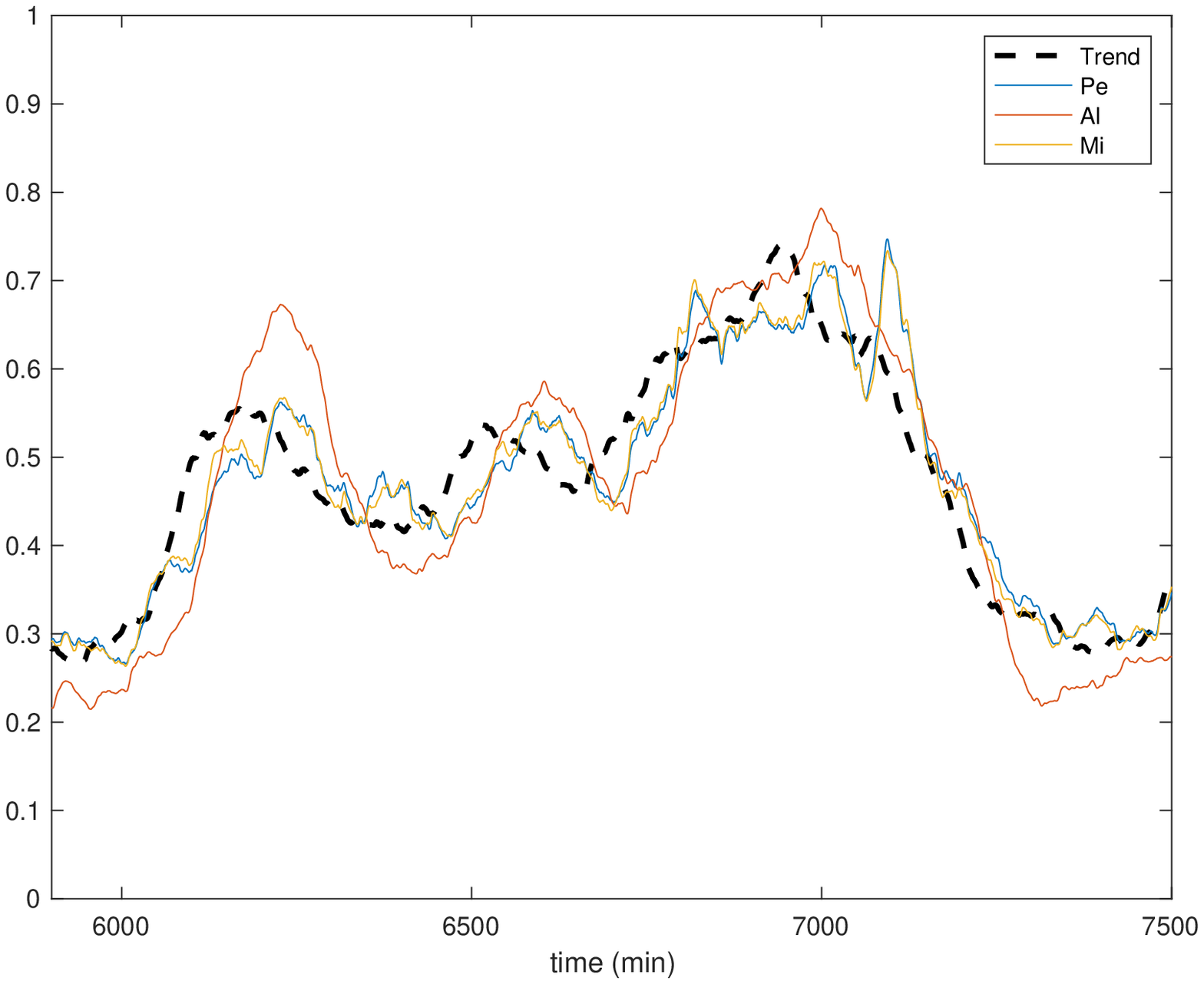,width=0.31\textwidth}}
\subfigure[\footnotesize Horizon: 60min]
{\epsfig{figure=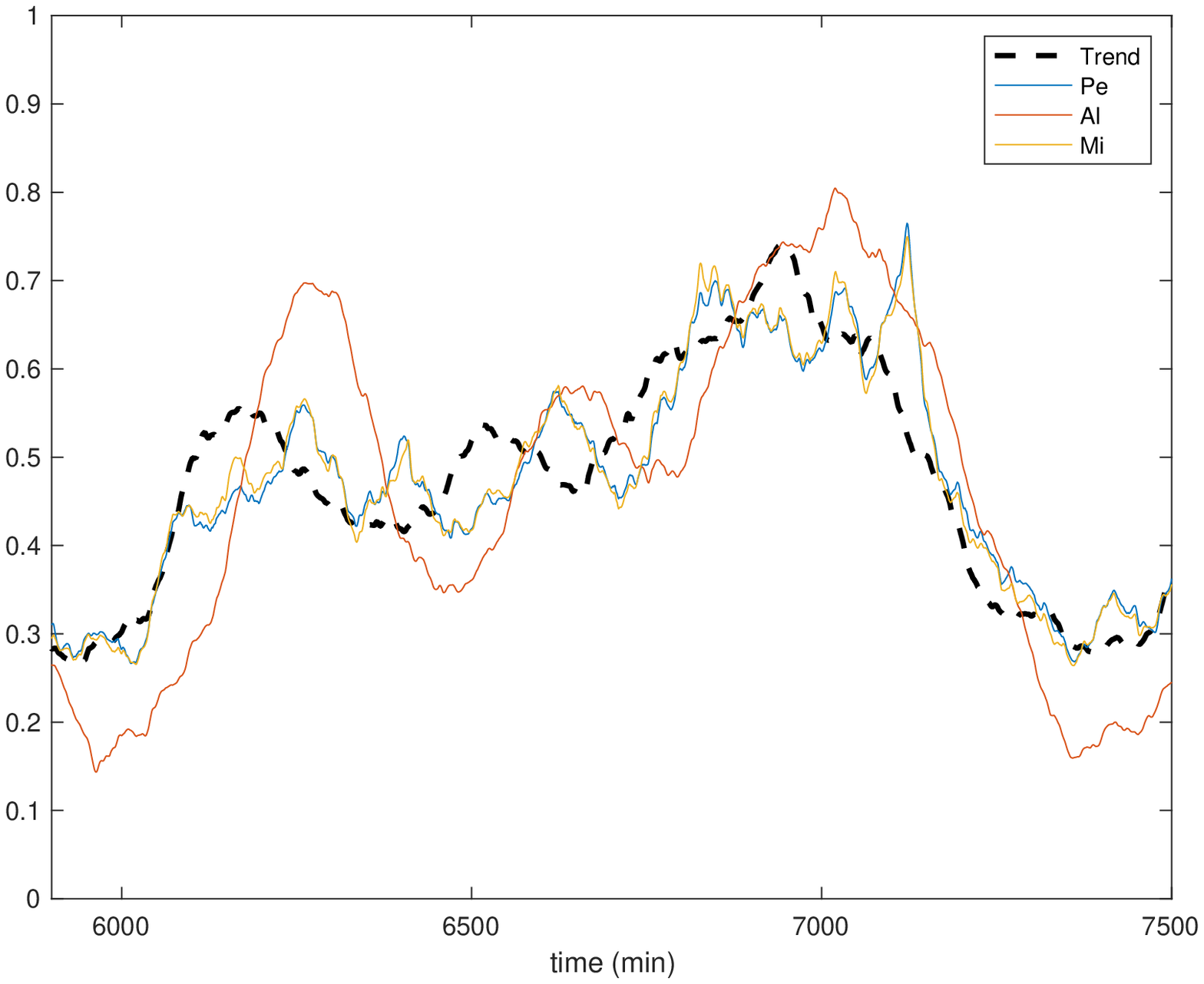,width=0.31\textwidth}}
\caption{Zoom2: Scaled persistence (\textcolor{blue}{\textbf{---}}), Algebraic techniques (\textcolor{red}{\textbf{---}}) and Algebraic techniques \& seasonality (\textcolor{yellow}{\textbf{---}}).}\label{Z2}
\end{figure*}


\begin{figure*}[!ht]
\centering
\subfigure[\footnotesize Horizon: 5min]
{\epsfig{figure=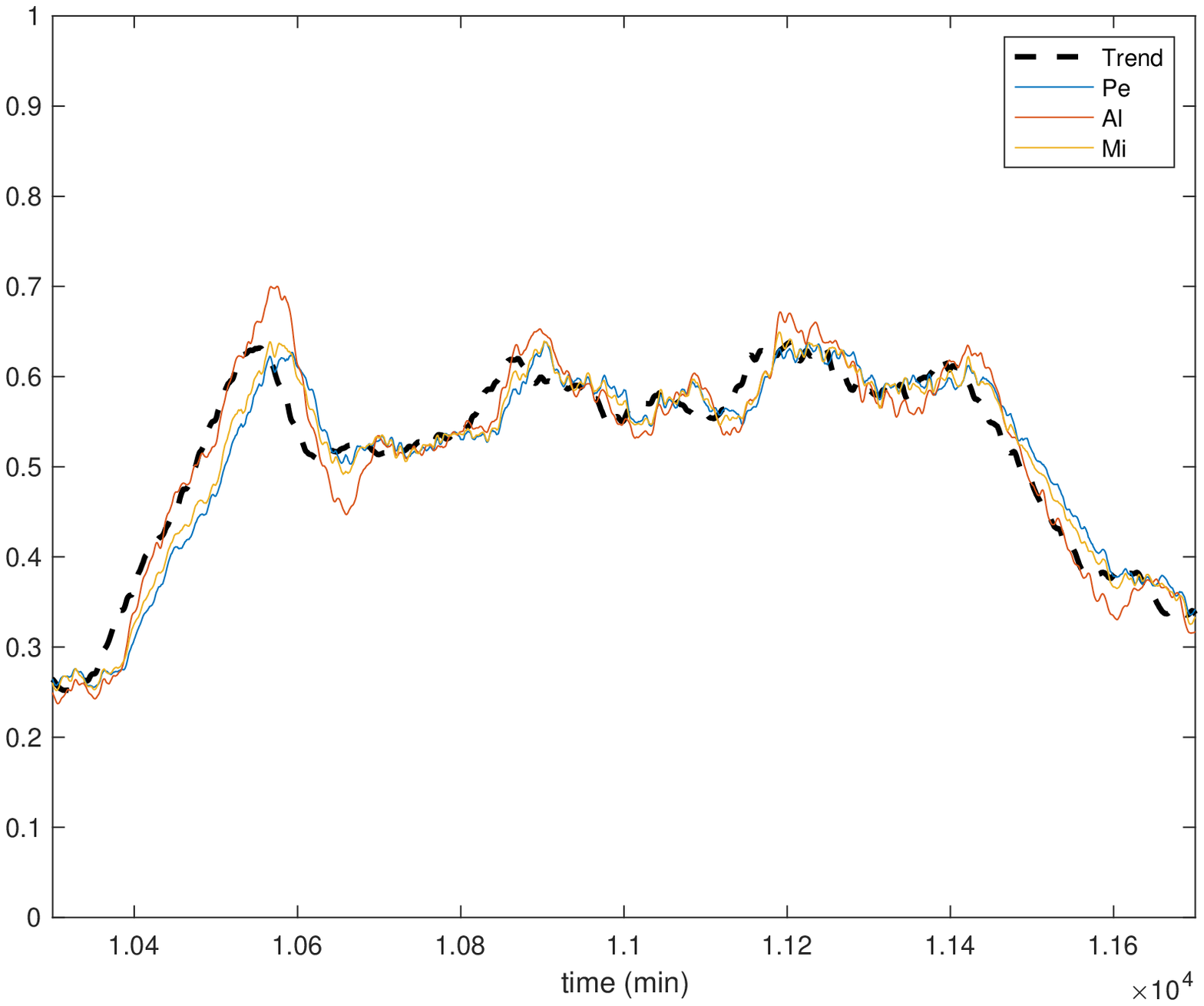,width=0.31\textwidth}}
\subfigure[\footnotesize Horizon: 30min]
{\epsfig{figure=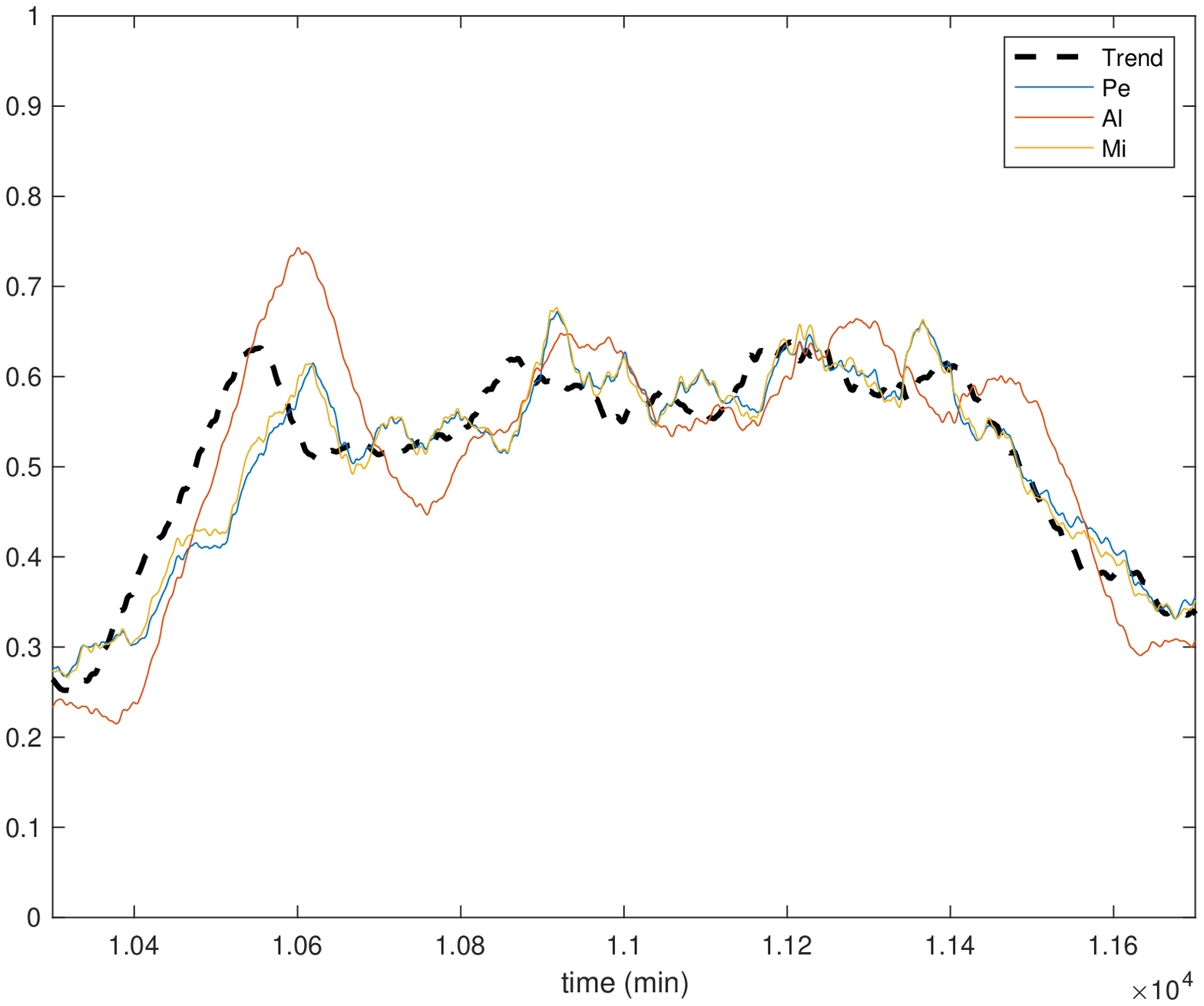,width=0.31\textwidth}}
\subfigure[\footnotesize Horizon: 60min]
{\epsfig{figure=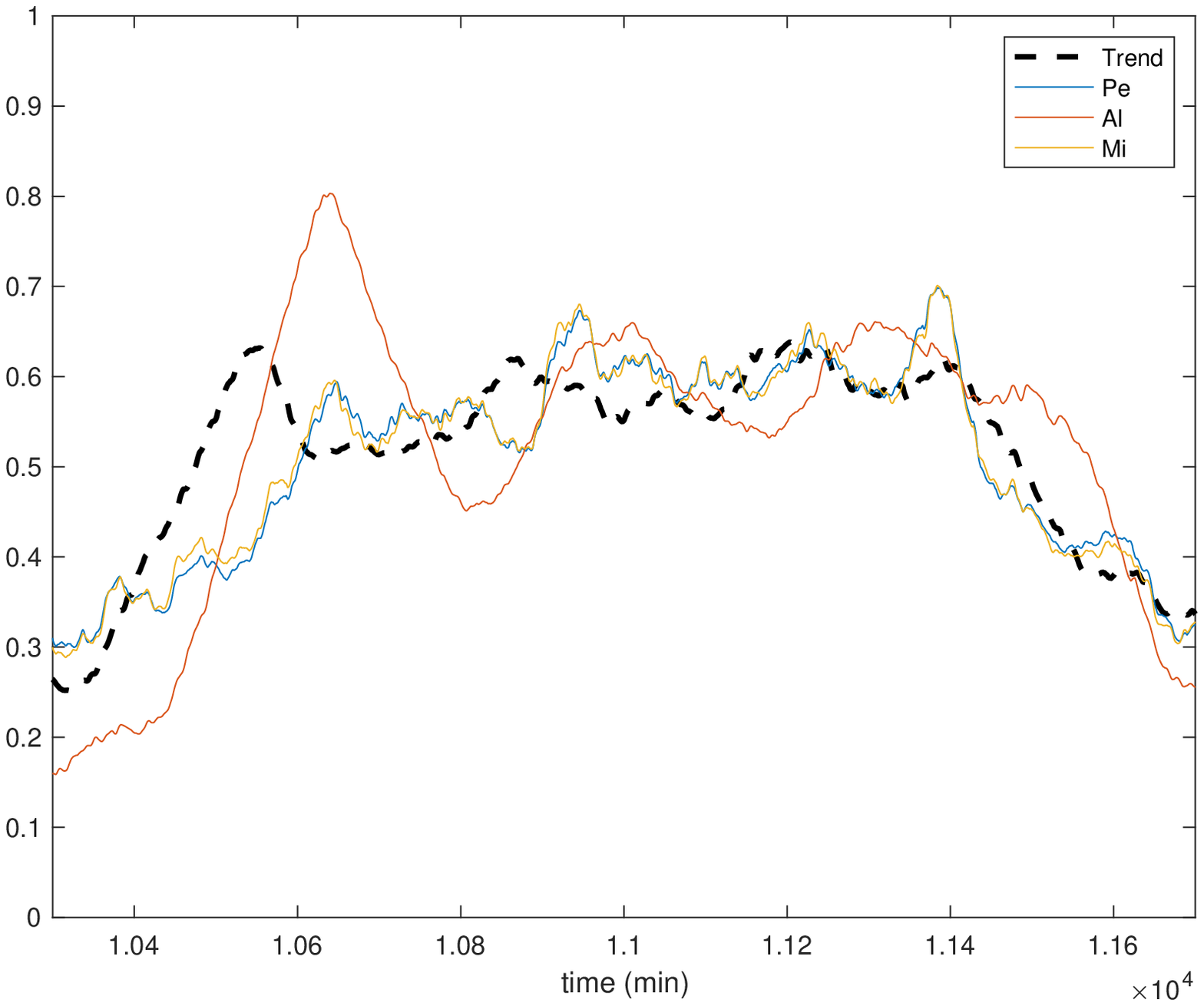,width=0.31\textwidth}}
\caption{Zoom3: Scaled persistence (\textcolor{blue}{\textbf{---}}), Algebraic techniques (\textcolor{red}{\textbf{---}}) and Algebraic techniques \& seasonality (\textcolor{yellow}{\textbf{---}}).}\label{Z3}
\end{figure*}

\section{Forecasting computing capacity needs for resource provisioning}\label{virtual}
Set $z(t)=5 \times 10^6 y(t)$: it is the processing time in milliseconds defined in Section \ref{A}.\footnote{The multiplicative factor $5 \times 10^6$ is "inspired" by the real data which are not made public in this paper.} Lack of space compels to a single forecast horizon of $30$ minutes via the single Formula \eqref{Mix}. See Figure \ref{WL} for $\widehat{z_{\rm trend}}(t+30)$.

In our case, the QoS and the Web Service stability are ensured with 50\% of processing usage on each single virtual CPU (vCPU) core. It yields the following predicted number of virtual machines: 
\begin{equation}\label{vm}
\widehat{n_{\rm VM}}(t+30{\rm min})=\frac{\widehat{z_{\rm trend}}(t+30{\rm min})}{30 000}
\end{equation}
where 
\begin{itemize}
\item \emph{VM} is the well-known acronym of \emph{virtual machine},
\item $30000$ is the number of milliseconds in a minute (in CPU time) for achieving the  $50\%$ use.
\end{itemize}
The computer results derived from Formula \eqref{vm} are provided in Figure \ref{VM}.
\begin{remark}
Being able to estimate the required amount of computing resources in advance, one can easily acquire the necessary resources via Cloud provider's spare available capacity at a bargained price. For example, in the case of \emph{Amazon Web Services} (\emph{AWS}), the use of Spot Instances can reduce the compute price up to 90\%.\footnote{AWS EC2 Spot: {\tt https$:$//aws.amazon.com/ec2/spot} }
\end{remark}

\begin{remark}
As already stated in our Introduction, the unavoidable quick oscillations around $\widehat{n_{\rm VM}}(t)$ and occasional uncertain load fluctuations are most efficiently taken into account \cite{iste} by iPs from model-free control.
\end{remark}



}

\begin{figure*}[!ht]
\centering
{\epsfig{figure=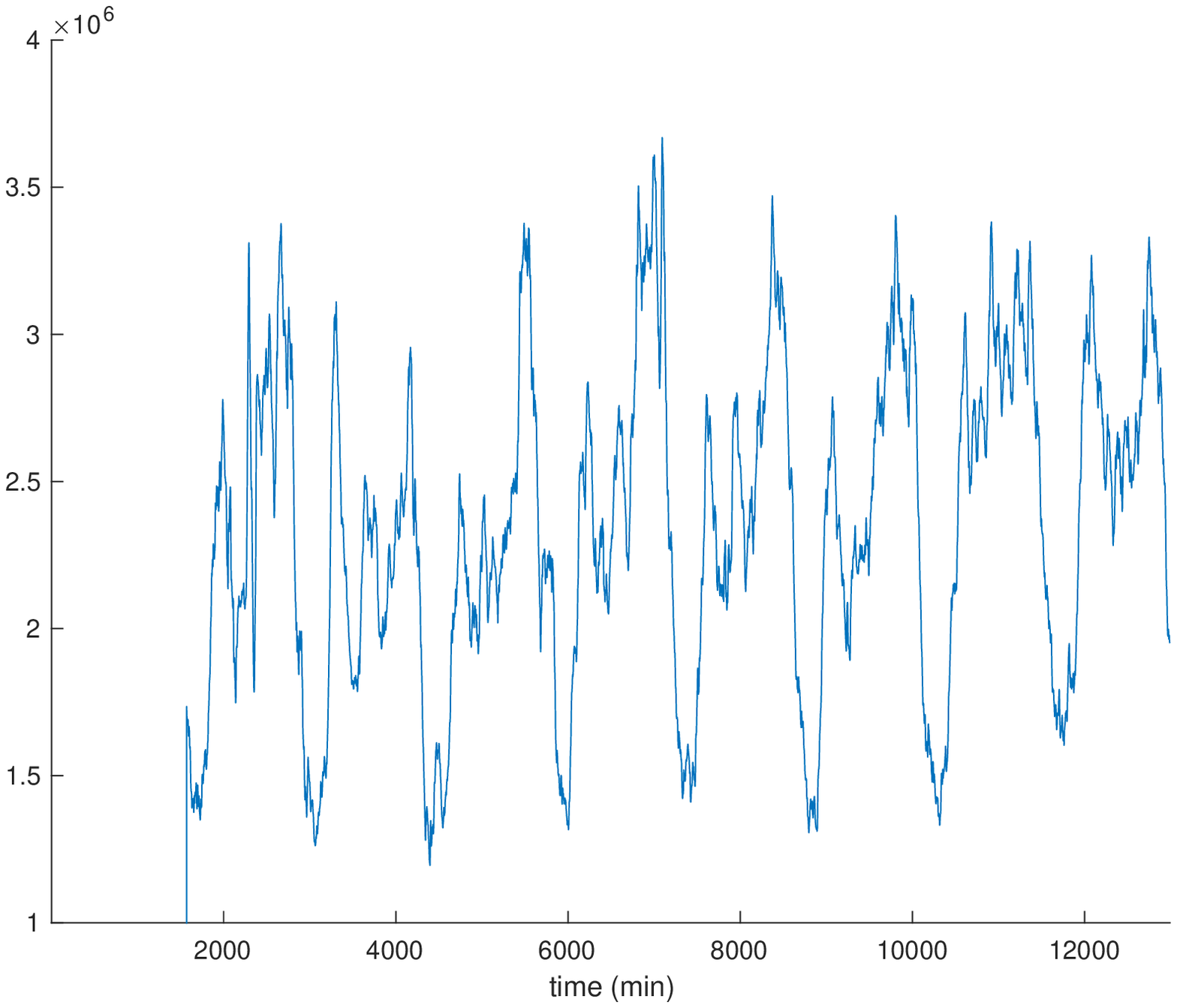,width=0.72\textwidth}}
\caption{Predicted workload via Section \ref{mixte}: $\widehat{z_{\rm trend}}(t+30{\rm min})$}\label{WL}
\end{figure*}

\begin{figure*}[!ht]
\centering
\subfigure[\footnotesize Zoom1]
{\epsfig{figure=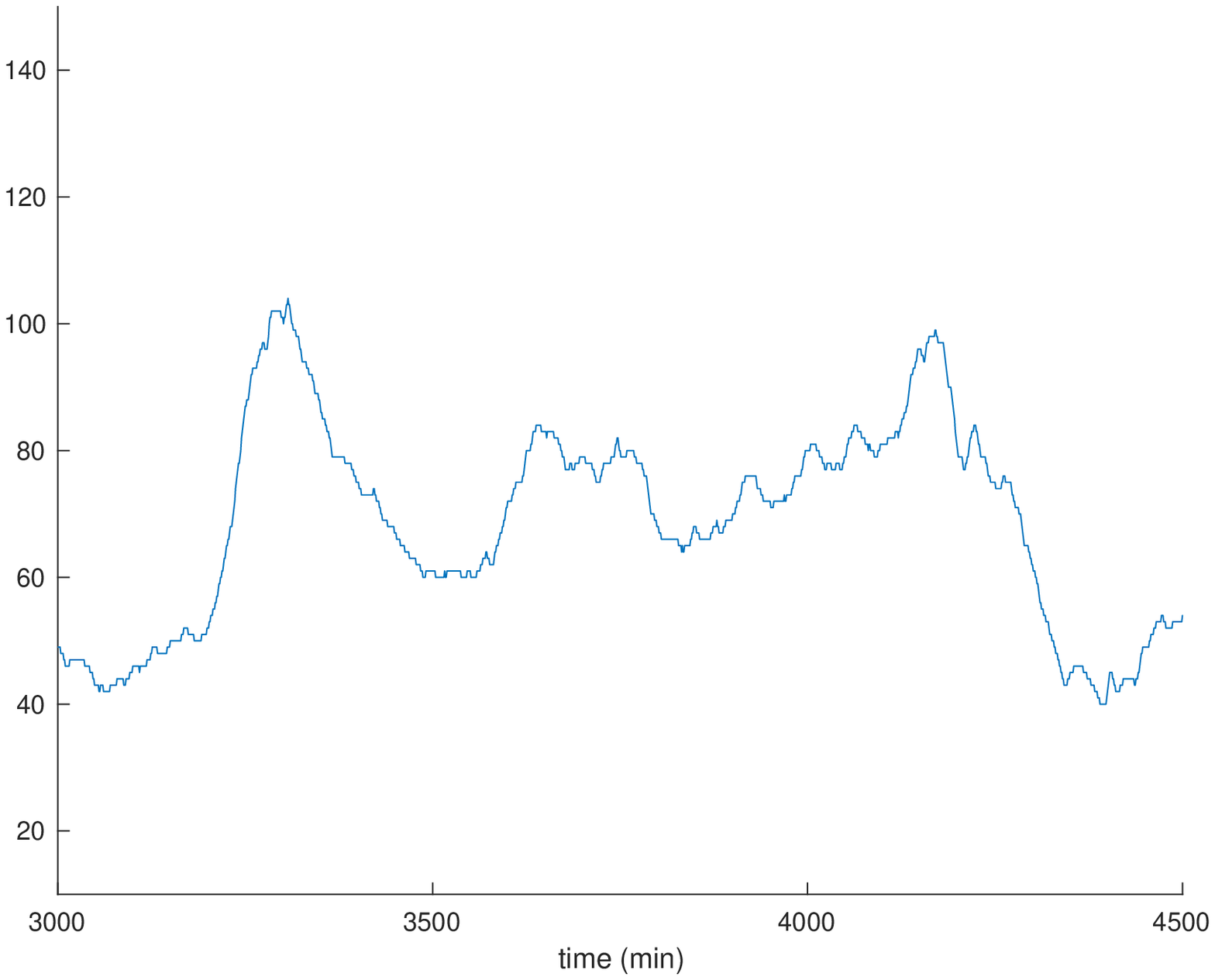,width=0.31\textwidth}}
\subfigure[\footnotesize Zoom2]
{\epsfig{figure=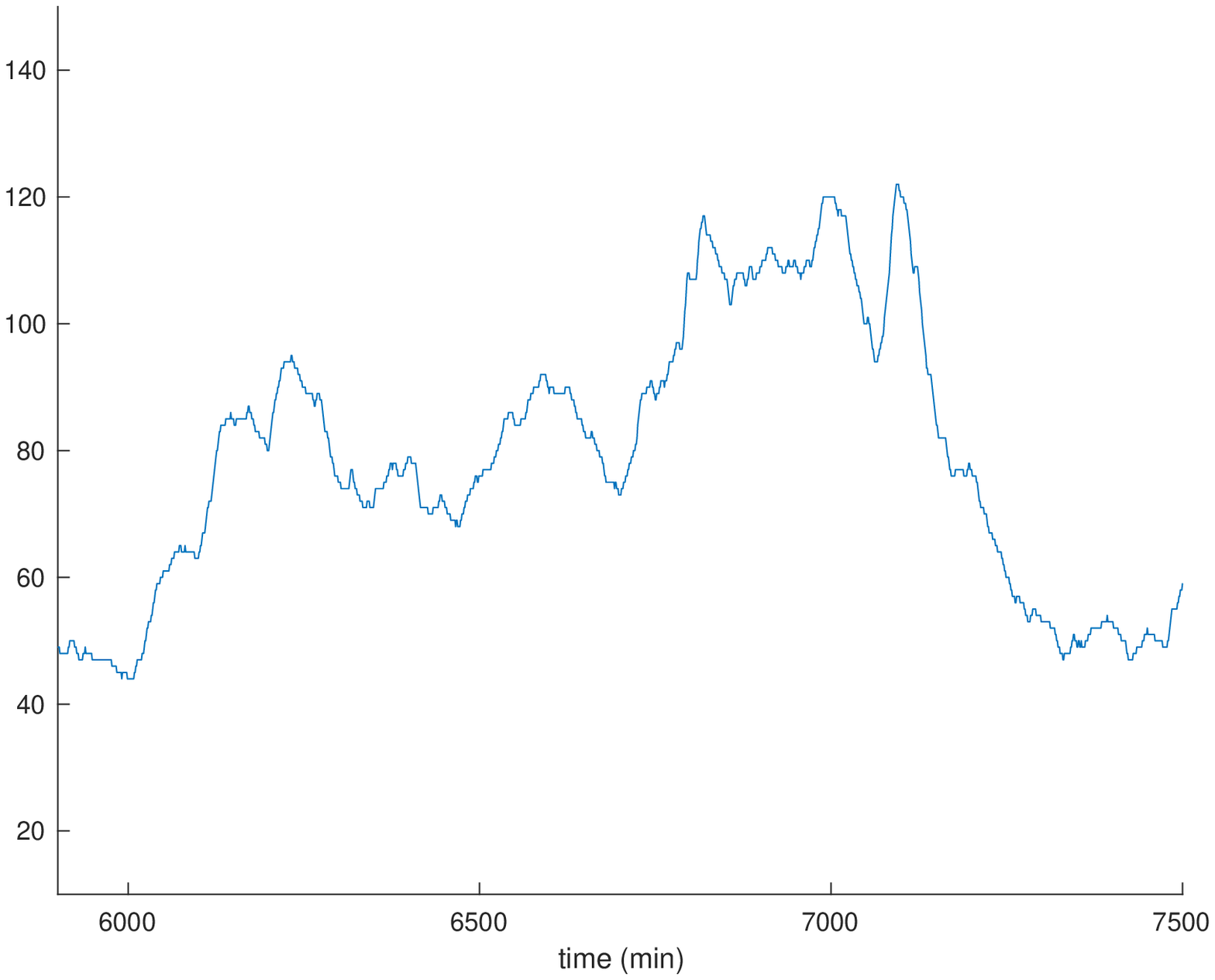,width=0.31\textwidth}}
\subfigure[\footnotesize Zoom3]
{\epsfig{figure=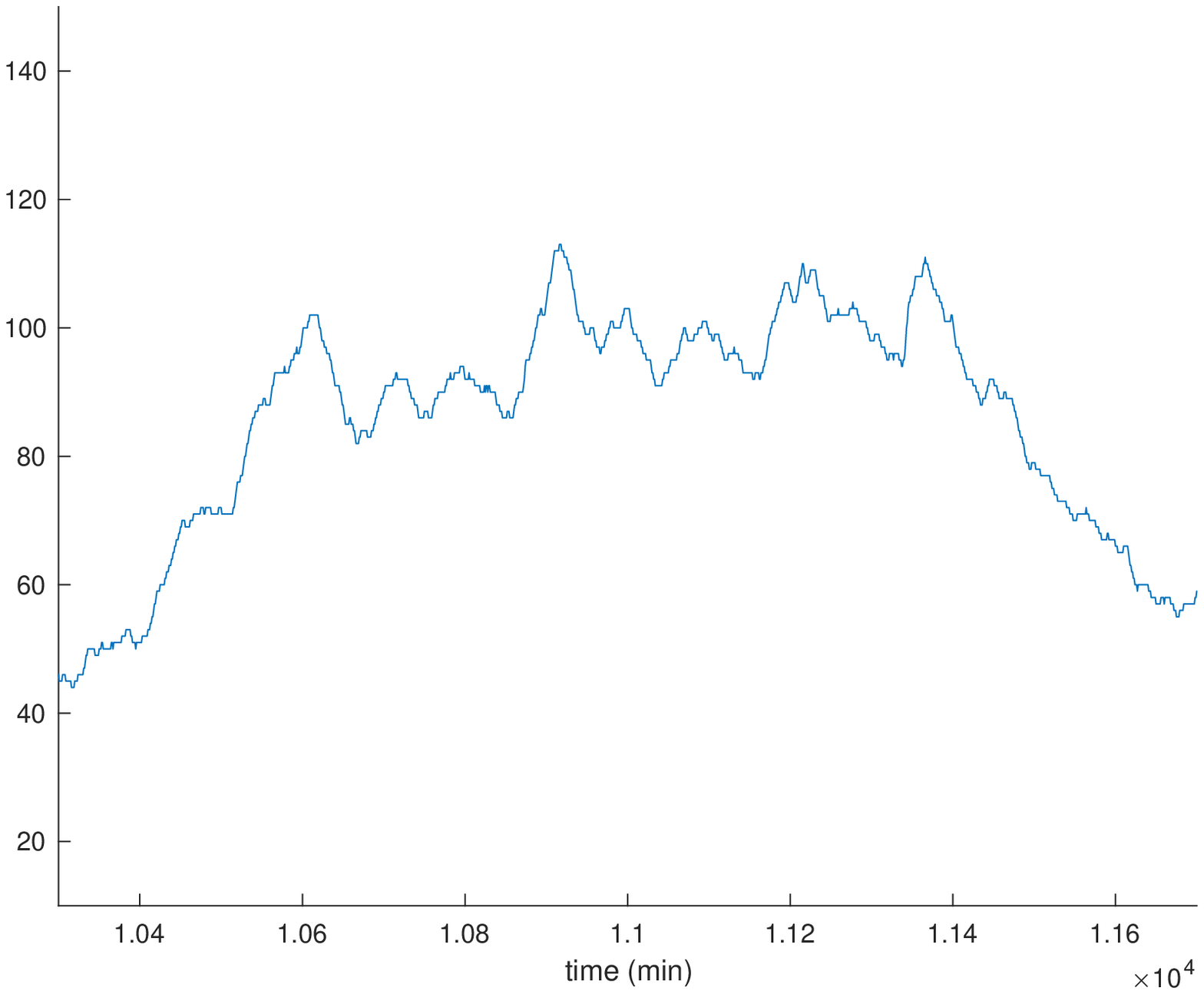,width=0.31\textwidth}}
\caption{Predicted need of virtual machines via Section \ref{mixte}: $\widehat{n_{\rm VM}}(t+30{\rm min})$}\label{VM}
\end{figure*}

\section{Conclusion}\label{conclusion}
We hope to have convinced the reader that easily implementable time series approaches yield convincing prediction results in cloud computing. If seasonality patterns are available, they should be exploited. They lead to notable simplifications and improvements.

This introductory communication should nevertheless be completed in several ways:
\begin{itemize}
\item Comparisons with other viewpoints on time series and, more generally, on forecasting should be investigated. Suitable metrics will be proposed.
\item How would behave our setting with respect to different time lapses and samplings?
\item The inevitable uncertainty of any forecasting technique plays obviously a critical role. It will be examined as in \cite{solar}.
\end{itemize}
Let us conclude by mentioning that some works have suggested to use time series for detecting \emph{anomalies} (see, \textit{e.g.}, \cite{huang2,vallis}) and enhencing \emph{privacy protection} (see, \textit{e.g.}, \cite{pawar,priv}). Our algebraic techniques \cite{rupt,solar} might also be helpful there.

\end{document}